\documentclass[aps,prd,onecolumn,showkeys,groupedaddress,showpacs,nofootinbib]{revtex4}
\usepackage{epsfig} \usepackage{amsmath} \usepackage{amsfonts}
\usepackage{amssymb} \usepackage{graphicx} \usepackage{colordvi}
\usepackage{psfrag}
 \usepackage{times}
 \usepackage{amsmath, amsthm, amssymb}
 \usepackage{makeidx}

\newtheorem{Proposition}{Proposition}[section]

\newfont{\gotico}{eufm10 scaled\magstephalf}
\newfont{\qvd}{msam10 scaled\magstephalf}

\def\demo{\par\noindent{\sc Proof. }\begingroup}
\def\enddemo{\hskip1em \mbox{\qvd \char3}\endgroup\par\medskip}

\def\de#1/de#2{\frac{\partial {#1}}{\partial {#2}}}
\def\De#1/de#2{\dfrac{\partial {#1}}{\partial {#2}}}

\def\const{{\rm const.}}

\makeindex \makeatletter
\def\widebar{\accentset{{\cc@style\underline{\mskip10mu}}}}
\makeatother

\begin{document}
%\numberwithin{equation}{section}
\def\bib#1{[{\ref{#1}}]}
\title{\bf On the well formulation of the Initial Value Problem of metric--affine $f(R)$-gravity}

\author{S. Capozziello$^{1}$ and S. Vignolo$^{2}$ }

\affiliation{$~^{1}$ Dipartimento di Scienze Fisiche,
Universit\`{a} ``Federico II'' di Napoli and INFN Sez. di Napoli,
Compl. Univ. Monte S. Angelo Ed. N, via Cinthia, I- 80126 Napoli
(Italy)}

\affiliation{$^{2}$DIPTEM Sez. Metodi e Modelli Matematici,
Universit\`a di Genova,  Piazzale Kennedy, Pad. D - 16129 Genova
(Italy)}

\date{\today}

\begin{abstract}
We study the well formulation of the initial value problem of
$f(R)$-gravity in the metric-affine formalism. The problem is
discussed in vacuo and in presence of perfect-fluid matter,
Klein-Gordon and Yang-Mills fields. Adopting Gaussian normal
coordinates, it can be shown that the problem is always {\it
well-formulated}. Our results refute some criticisms to the
viability of $f(R)$-gravity recently appeared in literature.
\end{abstract}


\maketitle

\section{Introduction}
Every theory of physics is "physically" viable if an appropriate
initial value problem is suitably formulated. This means that, starting from 
the assignment of suitable initial data, the
subsequent dynamical evolution of the physical system is uniquely
determined. In this case,
the problem is said {\it well-formulated}. For example, in
classical mechanics, given the initial positions and velocities of
the particles (or of the constituents of a physical system), if
the system evolves without external interferences, the dynamical
evolution is determined. This is true also for  field theories as,
for example, the Electromagnetism. However, also if the initial
value problem is  well-formulated, we need other further
properties that a viable theory has to satisfy. First of all,
small changes and perturbations in the initial data have to
produce small perturbations in the subsequent dynamics over all
the space-time where it is defined. This means that the theory
should be "stable" in order to be "predictive". Besides, changes
in the initial data region have to preserve the causal structure
of the theory. If both these requirements are satisfied, the
initial value problem of the theory is also {\it well-posed}.

It can be shown that General Relativity has a well-formulated and
well-posed initial value problem  but, as for other relativistic
field theories, we need initial value constraints and  gauge
choices in order to make Einstein's field equations suitable to
correctly formulate the Cauchy problem. The consequence of this
well-posedness is that General Relativity is a "stable" theory
with a robust causal structure where singularities can be
classified. A detailed discussion of these topics can be found in
 \cite{Synge,Wald}.

In this paper, we focus attention on the well formulation of the initial value
problem of metric--affine $f(R)$-theories of gravity, where $f(R)$ is a
generic function of the Ricci scalar $R$. The aim is to prove that the Cauchy
problem is well formulated in different and physically important cases.

The possible modifications of Einstein's theory  has a long
history which reaches back to the early times of General
Relativity
\cite{Weyl:1918,Pauli:1919,Weyl:1921,Bach:1921,Eddington:1924,Lanczos:1931}.
The early extensions and modifications  were aimed at a
unification of gravity with other fundamental interactions (in
particular the Electromagnetism). The recent interest in such
modifications comes from cosmological observations. For
comprehensive reviews, see for example
\cite{Schmidt:2004,OdiRev,GRGrew,faraonirev}. These observations
usually lead to the introduction of additional ad-hoc concepts
like dark energy/matter, if interpreted within Einstein's theory.
On the other hand, the emergence of such stopgap measures in a
cosmological context could be interpreted as a first signal of a
breakdown of General Relativity on these scales \cite{JCAP,MNRAS},
and led to the proposal of many alternative modifications of the
underlying gravity theory. In particular, more recent works
focused on the cosmological implications of $f(R)$-gravity since
such models may lead to an alternative explanation of the
acceleration effect observed in  cosmology
\cite{Capozziello:2002,Nojiri,Vollick:2003,Carroll:2004,Carroll:2005}
and to the explanation of the missing matter puzzle observed at
astrophysical scales \cite{MNRAS,salzano}.

While it is very natural to extend Einstein's gravity to theories
with additional geometric degrees of freedom, see for example
\cite{Hehl:1976,Hehl:1995,Trautman:2006} for some general surveys
on this subject as well as \cite{Puetzfeld:2005} for a list of
works in a cosmological context, recent attempts focused on the
old idea of modifying the gravitational Lagrangian in a purely
metric framework, leading to higher-order field equations.  Due to
the increased complexity of the field equations in this framework,
the main body of works dealt with some formally equivalent
theories, in which a reduction of the order of the field equations
was achieved by considering the metric and the connection as
independent objects \cite{francaviglia}. In addition, many authors
exploited the formal relationship to scalar-tensor theories to
make some statements about the weak field regime, which was
already worked out for scalar-tensor theories
\cite{Damour:Esposito-Farese:1992}.

However, a  concern which comes with generic higher-order gravity
theory is linked to the initial value problem. It is, up to now,
unclear if the prolongation of standard methods can be used in
order to tackle this problem in every theory. Hence it is doubtful
that the full Cauchy problem can be properly addressed, if one
takes into account the results already obtained in fourth-order
theories stemming from a quadratic Lagrangian
\cite{Teyssandier:Tourrenc:1983,Kerner}.

On the other hand, being $f(R)$-gravity, like General Relativity,
a gauge theory, the initial value formulation depends on suitable
constraints and on suitable "gauge choices" that mean a choice of
coordinates so that the Cauchy problem results well-formulated and
possibly well-posed. In \cite{Teyssandier:Tourrenc:1983,Noakes},
the initial value problem was studied for quadratic Lagrangians in
the metric approach with the conclusion that it is well-posed. On
the other hand, in \cite{Faraoni}, the Cauchy problem for generic
$f(R)$-models has been studied in metric and Palatini approaches
using the dynamical equivalence between these theories and the
Brans-Dicke gravity. The result was that the problem is
well-formulated for metric approach in presence of matter and
well-posed in vacuo. For the Palatini approach, instead, the
Chauchy problem is not well-formulated even in vacuo since,
considering the $3+1$ ADM formulation of the equivalent
scalar-tensor theory, the equivalent Brans-Dicke parameter
$\omega=-3/2$ leads to a non-dynamical equivalent field $\phi$ and
to the impossibility of a first-order formulation of the problem.

In this paper, we will adopt a different approach by which it is
possible to show that the Cauchy problem of metric-affine
$f(R)$-gravity is well-formulated and well-posed in vacuo, while
it can be, at least, well-formulated for various form of matter
fields like  perfect-fluids, Klein-Gordon and  Yang-Mills fields.
Therefore, our results refute the conclusions stated in \cite{Faraoni} on the
viability of metric-affine $f(R)$-theories of gravity. The reason
of the apparent contradiction with respect to the results in
\cite{Faraoni} lies on the above mentioned gauge choice. Following
\cite{Synge}, we adopt Gaussian normal coordinates. Such a choice,
introducing further constraints on the Cauchy data surface,
results more suitable to set the initial value problem in such a
way that the well-formulation is always possible. Of course, in order to
prove the complete viability of metric--affine $f(R)$-theories of gravity, 
the well posedness of the Cauchy problem has to be shown too. This 
topic is not dealt with in the present paper, being still an open problem 
under investigation. Anyway, here the important point is that, up to now and in our opinion, 
no objections to the viability of metric--affine $f(R)$-gravity based on the ill--formulation 
of the initial value problem can be made.

The layout of the paper is the following. In Sec.II, following \cite{Synge},
the initial value formulation of General Relativity is recalled.
Sec. III is devoted to set the problem  for $f(R)$-gravity in
metric-affine formalism in vacuo. In this case, the problem is
well-formulated and well-posed since dynamics reduces to the
Einstein theory plus cosmological constant. $f(R)$-gravity in
presence of matter is discussed in Sec.IV. As paradigmatic examples we discuss the coupling
with perfect-fluid matter and Yang-Mills fields, showing that in both the cases
the initial value problem results well-formulated.

In Sec.V, the
problem in presence of a Klein-Gordon scalar field is discussed.
In all these cases, the Gaussian normal coordinates allow the
well-formulation of the initial value problem. It is important to
stress that further constraint equations can emerge on the initial
surface and this fact could reduce the set of admissible initial
data.  Discussion and conclusions are given in Sec.VI. The
Appendix A is devoted to the demonstration of two useful
propositions related to the effective stress-energy tensor and the
Bianchi Identities.

\section{ Well formulation of the Cauchy problem for General Relativity}
Before starting with our considerations for $f(R)$-gravity, let us
recall the initial value formulation of General Relativity (i.e.
$f(R)=R$) where it is {\it well-formulated} (and also {\it
well-posed} as shown in \cite{Wald}). We adopt the formalism
developed in \cite{Synge}.

Let us consider a system of Gaussian normal coordinates
\cite{Wald} where the Latin indexes $i,j$ run from $1$ to $4$ and
the Greek indexes $\alpha$ run from $1$ to $3$. In these
coordinates, the time components of metric tensor are $g_{44}=-1$
and $g_{4\alpha}=0$ with the signature $(+\,+\,+\,-)$. These are
particularly useful to split the spatial hypersurface $S_3$ from
the orthogonal time-geodesics  in a given space-time $M$.

After, given a second rank symmetric tensor   $W_{ij}$, defined on
the globally hyperbolic space-time manifold $(M,g_{ij})$, it is
possible to define the symmetric conjugate tensor $W^*_{ij}$ as
$W^*_{ij}=W_{ij} -\frac{1}{2}Wg_{ij}$, where $W:=W^{ij}g_{ij}$ is
the trace of $W_{ij}$. Furthermore, if $S_4$ is a space-time
domain in $M$ where $g_{44}\not =0 $ and  $S_3$ is the $3$-surface
given by the equation $x^4=0$, then the following  statements are
equivalent:
\begin{enumerate}
    \item $W_{ij}=0$ in $S_4$.
    \item $W^*_{\alpha\beta}=0$ and $W_{4j}=0$ in $S_4$.
    \item $W^*_{\alpha\beta}=0$ and $W^i_{j|i}=0$ in $S_4$ with $W_{4j}=0$ in $S_3$.
\end{enumerate}
Let us take into account now the Einstein equations in the form
\begin{equation}\label{A.1}
G_{ij}=-kT_{ij} \quad\quad \mbox{with the Bianchi indentities}
\quad\quad T^{ij}_{\;\;|j}=0\,,
\end{equation}
where ${\displaystyle G_{ij}=R_{ij}-\frac{1}{2}g_{ij}R}$  is the
Einstein tensor. We can define the tensor
\begin{equation}\label{A.2}
W_{ij}:= G_{ij} + kT_{ij}\,.
\end{equation}
The conjugate tensor is
\begin{equation}\label{A.3}
W^*_{ij}=R_{ij} + kT^*_{ij}\,,
\end{equation}
and then the Einstein equations becomes
\begin{equation}\label{A.4}
W_{ij}=0\,.
\end{equation}
These are $10$ equations for  $20$ unknown functions $g_{ij}$ and
$T_{ij}$. Let us assign now the $10$ functions $g_{i4}$ and
$T_{\alpha\beta}$. The remaining $10$ functions $g_{\alpha\beta}$
and  $T_{i4}$, can be determined by Eqs. \eqref{A.4}. Using the
above results, these functions can be expressed in the equivalent
form
\begin{equation}\label{A.5}
R_{\alpha\beta} + kT^*_{\alpha\beta}=0\,, \qquad\qquad
W^i_{j|i}=T^i_{j|i}=0\,,
\end{equation}
with the  condition
\begin{equation}\label{A.6}
G_{4j} + kT_{4j}=0 \qquad \mbox{on the surface} \qquad x^4=0\,.
\end{equation}
Eqs. \eqref{A.5} can be rewritten in the form
\begin{subequations}\label{A.7}
\begin{equation}\label{A.7a}
g_{\alpha\beta,44}= 2\bar{R}_{\alpha\beta} -
\frac{1}{2}Ag_{\alpha\beta,4} +
g^{\mu\nu}g_{\alpha\mu,4}g_{\beta\nu,4} + 2kT^*_{\alpha\beta}\,,
\end{equation}
\begin{equation}\label{A.7b}
T_{4j,4}= - T^4_{j,4}=T^\alpha_{j,\alpha} +
\Gamma_{i\alpha}^{\;\;\;i}T^\alpha_j -
\Gamma_{ij}^{\;\;\;\alpha}T^i_\alpha\,,
\end{equation}
\end{subequations}
where $\bar{R}_{\alpha\beta}$ is the intrinsic  Ricci tensor
defined on the surface $x^4=0$, $\Gamma_{ij}^{\;\;\;k}$ is the
Levi-Civita connection related to the metric $g_{ij}$, the coma is
the partial derivative and
\begin{equation}\label{A.8}
A:=g^{\mu\nu}g_{\mu\nu,4}\,.
\end{equation}
In the same way, the constraint equations \eqref{A.6} become
\begin{subequations}\label{A.9}
\begin{equation}\label{A.9a}
A_{,\alpha} - D^{\sigma}g_{\alpha\sigma,4} + 2kT_{4\alpha}=0\,,
\end{equation}
\begin{equation}\label{A.9b}
\bar{R} - \frac{1}{4}A^2 + \frac{1}{4}B + 2kT_{44}=0\,,
\end{equation}
\end{subequations}
where $\bar{R}$ is the intrinsic curvature scalar  of  the surface
$x^4=0$, $D_\sigma$ denotes the covariant derivatives on the
surface $x^4=0$ associated to the Levi-Civita connection of the
intrinsic metric ${g_{\alpha\beta}}_{|x^4=0}$ and
\begin{equation}\label{A.10}
B=g^{\mu\nu}g^{\rho\sigma}g_{\mu\rho,4}g_{\nu\sigma,4}\,.
\end{equation}
Let us assign now the set of Cauchy data on the surface $x^4=0$
\begin{equation}\label{A.11}
g_{\alpha\beta}, \quad g_{\alpha\beta,4}, \quad T_{i4}\,.
\end{equation}
Such data have to satisfy the constraint equations \eqref{A.9}.
Eqs. \eqref{A.7} explicitly give the values of the quantities
\begin{equation}\label{A.12}
g_{\alpha\beta,44}, \qquad T_{4j,4}\,,
\end{equation}
as a function of the Cauchy data. By deriving Eqs. \eqref{A.7}, it
is straightforward to obtain the time-derivatives of further order
as a function of the Cauchy data. This procedure allows to
reconstruct the solution of the field equations as a power-law
series of the time variable $x^4$.

In other words, this means that the 3-surface $S_3$, given by the
equation  $x^4=0$, is a Cauchy surface of the globally hyperbolic
space-time $(M,g_{ij})$ and that the initial value formulation
(the Cauchy problem) is well-formulated in General
Relativity. Our task is now to extend these results to
$f(R)$-gravity in metric-affine formalism.

\section{The Cauchy problem for $f(R)$-gravity in metric-affine formalism in empty space }
In the metric-affine formulation of $f(R)$-gravity, the dynamical
fields are given by the couple of functions $(g,\Gamma)\/$ where
$g\/$ is the metric and  $\Gamma\/$ is the linear connection. In
vacuo, the field equations are obtained by varying with respect to
the metric and the connection the following action
\begin{equation}\label{2.0}
{\cal A}\/(g,\Gamma)=\int{\sqrt{|g|}f\/(R)\,ds}
\end{equation}
where $f(R)$ is a real function, $R\/(g,\Gamma) = g^{ij}R_{ij}\/$
(with $R_{ij}:= R^h_{\;\;ihj}\/$) is the scalar curvature
associated to the dynamical connection $\Gamma\/$.

More precisely, in the approach with torsion, one can ask for a
metric connection $\Gamma$ with torsion different from zero while,
in the Palatini approach, the $\Gamma$ is non-metric but torsion
is null \cite{CCSV1}.

In vacuo, the field equations for $f(R)$-gravity with torsion are
\cite{CCSV1,CCSV2,CCSV3}
\begin{subequations}\label{2.1}
\begin{equation}\label{2.1a}
f'\/(R)R_{ij} - \frac{1}{2}f\/(R)g_{ij}=0\,,
\end{equation}
\begin{equation}\label{2.1b}
T_{ij}^{\;\;\;h} = -
\frac{1}{2f'}\de{f'}/de{x^p}\/\left(\delta^p_i\delta^h_j -
\delta^p_j\delta^h_i\right)\,,
\end{equation}
\end{subequations}
while the field equations for $f(R)$-gravity  {\it \`{a} la}
Palatini are \cite{francaviglia,Sotiriou,Sotiriou-Liberati1,Olmo}
\begin{subequations}\label{2.2}
\begin{equation}\label{2.2a}
f'\/(R)R_{ij} - \frac{1}{2}f\/(R)g_{ij}=0\,,
\end{equation}
\begin{equation}\label{2.2b}
\nabla_k\/(f'(R)g_{ij})=0\,.
\end{equation}
\end{subequations}
In both cases, considering the trace of  Einstein-like field
equations  \eqref{2.1a} e \eqref{2.2a}, one gets
\begin{equation}\label{2.3}
f'\/(R)R  - 2f\/(R)=0\,.
\end{equation}
It is easy to conclude that scalar curvature $R$ is a constant
coinciding with the solution of Eq.\eqref{2.3}. In this case, Eqs.
\eqref{2.1b} e \eqref{2.2b} imply that both dynamical connections
 coincide with the  Levi--Civita connection  associated to the
metric $g_{ij}$ which is solution of the field equations. In other
words, both theories reduce to the Einstein theory plus
cosmological constant.  As it is well known, in this case the Cauchy problem is well
formulated and well posed too \cite{Wald}.

\section{The Cauchy problem in presence of  matter}
Let us now take into account the presence of perfect-fluid matter
in the formulation of the Cauchy problem for $f(R)$-gravity. In
order to deal simultaneously with the Palatini approach and the
torsion, we will consider the connection not coupled with matter.
In other words, we will assume that the matter Lagrangian does not
explicitly depend on the dynamical connection. With this working
hypothesis, the field equations are
\begin{subequations}\label{3.1}
\begin{equation}\label{3.1a}
f'\/(R)R_{ij} - \frac{1}{2}f\/(R)g_{ij}=\Sigma_{ij}\,,
\end{equation}
with
\begin{equation}\label{3.1b}
T_{ij}^{\;\;\;h} = - \frac{1}{2f'\/(R)}\de{f'\/(R)}/de{x^p}\/\left(\delta^p_i\delta^h_j - \delta^p_j\delta^h_i\right)
\end{equation}
\end{subequations}
in the case of $f(R)$-gravity with torsion, and
\begin{subequations}\label{3.2}
\begin{equation}\label{3.2a}
f'\/(R)R_{ij} - \frac{1}{2}f\/(R)g_{ij}=\Sigma_{ij}\,,
\end{equation}
\begin{equation}\label{3.2b}
\nabla_k\/(f'(R)g_{ij})=0\,,
\end{equation}
\end{subequations}
in the case of $f(R)$-gravity in the  Palatini approach. In Eqs.
\eqref{3.1a} and \eqref{3.2a}, ${\displaystyle \Sigma_{ij}:= -
\frac{1}{\sqrt{|g|}}\frac{\delta{\cal L}_m}{\delta g^{ij}}\/}$ is
the stress-energy tensor. Considering the trace of Eqs.
\eqref{3.1a} and \eqref{3.2a}, we obtain a relation between the
curvature scalar $R\/$ and the trace of the stress-energy tensor
$\Sigma:=g^{ij}\Sigma_{ij}\/$. We have
\begin{equation}\label{3.3}
f'\/(R)R -2f\/(R) = \Sigma\,.
\end{equation}
It is worth noticing that any time that $\Sigma=\const$ the theory
reduces to the General Relativity with cosmological constant and
the initial value problem is identical to the above empty-space
case.

By the hypotheses that the relation \eqref{3.3} is invertible and
$\Sigma \not= \const\/$, the curvature scalar  $R\/$ can be
expressed as a function of  $\Sigma\/$, that is
\begin{equation}\label{3.4}
R=F\/(\Sigma)\,.
\end{equation}
 Starting from this fact, it is easy to show
that the Einstein-like equations of both Palatini and
metric-affine theory with torsion can be expressed in the same
form \cite{CCSV1,CCSV2,Olmo}, that is
\begin{equation}\label{3.5}
\begin{split}
\tilde{R}_{ij} -\frac{1}{2}\tilde{R}g_{ij}= \frac{1}{\varphi}\Sigma_{ij}
+ \frac{1}{\varphi^2}\left( - \frac{3}{2}\de\varphi/de{x^i}\de\varphi/de{x^j}
+ \varphi\tilde{\nabla}_{j}\de\varphi/de{x^i} + \frac{3}{4}\de\varphi/de{x^h}\de\varphi/de{x^k}g^{hk}g_{ij} \right. \\
\left. - \varphi\tilde{\nabla}^h\de\varphi/de{x^h}g_{ij} -
V\/(\varphi)g_{ij} \right)\,,
\end{split}
\end{equation}
where we have defined the effective potential
\bigskip\noindent
\begin{equation}\label{3.6}
V\/(\varphi):= \frac{1}{4}\left[ \varphi
F^{-1}\/((f')^{-1}\/(\varphi)) +
\varphi^2\/(f')^{-1}\/(\varphi)\right]\,,
\end{equation}
for the scalar field
\begin{equation}\label{3.7}
\varphi := f'\/(F\/(\Sigma))\,.
\end{equation}
Introducing the conformal transformation
$\bar{g}_{ij}=\varphi\/g_{ij}$, Eq. \eqref{3.5} assume the simpler
form (see for example \cite{CCSV1,Olmo,German})
\begin{equation}\label{3.8}
\bar{R}_{ij} - \frac{1}{2}\bar{R}\bar{g}_{ij} =
\frac{1}{\varphi}\Sigma_{ij} -
\frac{1}{\varphi^3}V\/(\varphi)\bar{g}_{ij}\,,
\end{equation}
where $\bar{R}_{ij}\/$ and $\bar{R}\/$ are respectively the Ricci
tensor and the curvature scalar derived from the conformal metric
$\bar{g}_{ij}\/$. It is worth noticing that the conformal
transformation is working if the trace $\Sigma$ of the
stress-energy tensor is independent of the metric $g_{ij}$. In
such a case, Eqs.\eqref{3.8} depend exclusively on the conformal
metric $\bar{g}_{ij}$ and the other matter fields.

Furthermore, we have to stress that the connection $\Gamma$,
solution of the field equations with torsion, is given by
\begin{equation}\label{3.9}
\Gamma_{ij}^{\;\;\;h} =\tilde{\Gamma}_{ij}^{\;\;\;h} +
\frac{1}{2\varphi}\de\varphi/de{x^j}\delta^h_i -
\frac{1}{2\varphi}\de\varphi/de{x^p}g^{ph}g_{ij}\,,
\end{equation}
where $\tilde{\Gamma}_{ij}^{\;\;\;h}$ is the Levi-Civita
connection  induced by the metric $g_{ij}$, while the connection
$\bar\Gamma$,  solution of the Palatini field equations, coincides
with the Levi-Civita connection  associated to the conformal
metric  $\bar{g}_{ij}$. Both connections, $\Gamma$ and
$\bar\Gamma$, satisfy the relation
\begin{equation}\label{3.10}
\bar{\Gamma}_{ij}^{\;\;\;h} =\Gamma_{ij}^{\;\;\;h} +
\frac{1}{2\varphi}\de\varphi/de{x^i}\delta^h_j\,.
\end{equation}
The identities
\begin{equation}\label{3.11}
\bar{\Gamma}_{ij}^{\;\;\;h}= \tilde{\Gamma}_{ij}^{\;\;\;h} +
\frac{1}{2\varphi}\de\varphi/de{x^j}\delta^h_i -
\frac{1}{2\varphi}\de\varphi/de{x^p}g^{ph}g_{ij} +
\frac{1}{2\varphi}\de\varphi/de{x^i}\delta^h_j\,,
\end{equation}
hold. This result shows the relation between the Levi-Civita
connections  induced by the metrics $g_{ij}$ and $\bar{g}_{ij}$.

In general, the Einstein-like Eqs. \eqref{3.5} have to be
considered together with the matter field equations. To this
purpose, we have to keep in mind that the conservation equations
for both the metric-affine theories (with torsion and {\it \`{a}
la} Palatini) coincide with the standard conservation laws of
General Relativity (see the Appendix A).  This means
\begin{equation}\label{3.12} \tilde\nabla_j\Sigma^{ij}=0\,.
\end{equation}
It is straightforward to show that Eqs. \eqref{3.12} are
equivalent to the conservation laws
\begin{equation}\label{3.12a}
\bar{\nabla}_jT^{ij}=0 \qquad {\rm where} \qquad T_{ij}=
\frac{1}{\varphi}\Sigma_{ij} -
\frac{1}{\varphi^3}V\/(\varphi)\bar{g}_{ij}\,,
\end{equation}
for the conformally transformed theories \eqref{3.8}. In fact, by
an explicit calculation of the divergence  $\bar{\nabla}_jT^{ij}$
where the relations \eqref{3.11} have been used, we obtain the
equations
\begin{equation}\label{3.12b}
\bar{\nabla}^jT_{ij}= \frac{1}{\varphi^2}\tilde\nabla^j\Sigma_{ij}
+ \frac{1}{\varphi^3}\de\varphi/de{x^i}\left( -\frac{1}{2}\Sigma +
\frac{3}{\varphi}V(\varphi) - V'(\varphi)\right)\,.
\end{equation}

The constraint equations \eqref{3.12} and \eqref{3.12a} are then
mathematically equivalent in view  of the  relation
\begin{equation}\label{3.12c}
\Sigma -\frac{6}{\varphi}V(\varphi) + 2V'(\varphi)=0\,,
\end{equation}
which is equivalent to the definition $\varphi=f'(F(\Sigma))$
\cite{CCSV1}.

With these results in mind, the Cauchy problem for Eqs.
\eqref{3.5} and the related matter field equations can be faced by
discussing the equivalent initial value problem of the conformally
transformed theories. Considering as in the above General
Relativity case, Gaussian normal coordinates and starting  from
Eqs. \eqref{3.8} and \eqref{3.12a}, it is easy to conclude that
the Cauchy problem is well formulated also in this case.

It is worth pointing out that, in general,  the matter field
equations imply the  Levi-Civita connection  derived from the
metric $g_{ij}$ and not the connection induced from the conformal
metric $\bar{g}_{ij}$. In view of the relation \eqref{3.11}, this
is not a problem since the  connection $\tilde\Gamma$ can be
expressed as a function of the connection $\bar\Gamma$  and of the
 scalar field $\varphi$, which, on the other hand, is a function of the source matter fields.
As a result, we could obtain slightly more complicated equations
implying further constraints on the initial data but, in any case,
the same equations can be always recast in "normal form" with
respect to the maximal order time derivatives  of matter fields
allowing a well-formulated Cauchy problem \cite{Wald}.

As an example, let us examine in detail the perfect-fluid case
with equation of state $p=p(\rho)$. The corresponding
stress-energy tensor is
\begin{equation}\label{3.13}
\Sigma_{ij} = \rho\/V_iV_j + p\/\left(V_iV_j + g_{ij}\right)\,,
\end{equation}
the matter field equations  are given by Eqs. \eqref{3.12} with
the further condition
\begin{equation}\label{3.14}
g_{ij}V^iV^j =-1\,,
\end{equation}
where $V^j$ are 4-velocities. Specifically, Eqs. \eqref{3.12} give
the field equations
\begin{equation}\label{3.15.0}
\left(\rho + pV^j\right)_{|j}V_i + \left(\rho + p\right)V_{i|j}V^j
+ \de p/de{x^i}=0\,,
\end{equation}
where $|$ denotes the covariant derivative with respect to the
Levi-Civita connection induced by $g_{ij}$. By saturating with
$V^i$, one gets
\begin{subequations}\label{3.15}
\begin{equation}\label{3.15a}
\left(\rho\/V^j \right)_{|j} = -pV^j_{|j}\,,
\end{equation}
while by substituting Eq. \eqref{3.15a} into Eqs. \eqref{3.15.0}
for $\alpha=1,2,3$, we have
\begin{equation}\label{3.15b}
\left(\rho+p\right)V^jV^\alpha_{|j}= -\de
p/de{x^j}\left(V^{\alpha}V^j + g^{\alpha j}\right)\,.
\end{equation}
\end{subequations}
As already pointed out, Eqs. \eqref{3.14} and \eqref{3.15} involve
the metric $g_{ij}$ and its first derivatives. By using the
relations \eqref{3.11}, we can rewrite these equations in the
conformal metric $\bar{g}_{ij}$, the scalar field function (of the
matter density)  $\varphi=\varphi\/(\rho)$ and their first
derivatives. Immediately we get
\begin{subequations}\label{3.16}
\begin{equation}\label{3.16a}
\frac{1}{\varphi}\bar{g}_{ij}V^iV^j =-1\,,
\end{equation}
\begin{equation}\label{3.16b}
\de /de{x^j}\left(\rho\/V^j\right) +
\bar{\Gamma}_{js}^{\;\;\;j}\rho\/V^s
-\frac{2}{\varphi}\de\varphi/de{x^s}\rho\/V^s = -p\left(\de
V^j/de{x^j} + \bar{\Gamma}_{js}^{\;\;\;j}V^s -
\frac{2}{\varphi}\de\varphi/de{x^s}\/V^s\right)\,,
\end{equation}
\begin{equation}\label{3.16c}
\begin{split}
\left(\rho + p\right)V^j\left[\de V^\alpha/de{x^j} +
\bar{\Gamma}_{js}^{\;\;\;\alpha}\/V^s + \frac{1}{2\varphi}\left(-\de\varphi/de{x^s}\delta^\alpha_j +
\de\varphi/de{x^j}\delta^\alpha_s - \de\varphi/de{x^p}g^{p\alpha}g_{js}\right)V^s\right]\\
= - \de p/de{x^j}\left(V^{\alpha}V^j + g^{\alpha j}\right)\,.
\end{split}
\end{equation}
\end{subequations}
In the Gaussian normal coordinates, where $\bar{g}_{44}=\pm 1$
(depending on the sign of $\varphi$) and $\bar{g}_{\alpha 4}=0$,
we obtain, from Eq. \eqref{3.16a}, the expression of $V^4$ in
terms of the remaining $V^\alpha$ . Eqs. \eqref{3.16a} and
\eqref{3.16b} can be considered as linear equations for the
functions $\De{V^\alpha}/de{x^4}$ and $\De\rho/de{x^4}$. The
explicit resolution of these equations, in terms of the unknown
functions, could give rise to further constraints on the initial
data and on the form of the $f(R)$-function. In any case, from
Eqs. \eqref{3.16a} and \eqref{3.16b} and in Gaussian normal
coordinates, one can derive the quantities $\De{V^\alpha}/de{x^4}$
e $\De\rho/de{x^4}$ as functions of the initial data
$\bar{g}_{\alpha\beta}$, $\de{\bar{g}_{\alpha\beta}}/de{x^4}$,
$V_\alpha$ and $\rho$ allowing to put the matter-field equations
in normal form. This means that the Cauchy problem is
well-formulated.

As another important example, let us consider the initial value
formulation of $f(R)$-gravity coupled with Yang-Mills fields,  in
particular with Electromagnetic field. Also in this case, the
problem is well-formulated. In fact, the stress-energy tensor of a
Yang-Mills field has null trace. From Eq.\eqref{3.3}, it is easy
to prove that the  curvature scalar $R$ is  constant and then, by
using Eqs. \eqref{3.1b} and \eqref{3.2b}, it is straightforward to
conclude that the dynamical connection coincides with the
Levi-Civita connection of the metric $g_{ij}$. In this situation,
both theories ({\it \`{a} la} Palatini and with torsion) reduce to
the Einstein theory with cosmological constant and then the Cauchy
problem is well-formulated (this last result is immediate for the
Electromagnetic field). Regarding the well-posed initial
formulation, the above results work for any theory where the trace
of stress-energy tensor is a constant since, as above,  the
$f(R)$-gravity reduces to Einstein gravity plus cosmological
constant.

\section{The Cauchy problem in the case of coupling with a scalar field}

Let us consider now the case of coupling with a Klein-Gordon
scalar field $\psi$ with self-interacting potential
${\displaystyle U(\psi)=\frac{1}{2} m^2\psi^2} $. The
stress-energy tensor is given by
\begin{equation}\label{4.1}
\Sigma_{ij}= \de\psi/de{x^i}\de\psi/de{x^j}
-\frac{1}{2}g^{ij}\left(\de\psi/de{x^p}\de\psi/de{x^q}g^{pq} +
m^2\psi^2\right)\,.
\end{equation}
The trace of tensor \eqref{4.1} is
\begin{equation}\label{4.2}
\Sigma =-\de\psi/de{x^p}\de\psi/de{x^q}g^{pq} -2m^2\psi^2\,.
\end{equation}
In the relation \eqref{4.2}, the metric $g^{pq}$ explicitly
appears. This means that the conformal transformation procedure of
the previous section cannot be  applied in this case and then the
well-formulation of the Cauchy problem has to be directly shown
starting from the field Eqs. \eqref{3.5} and the  Klein-Gordon
equation (see also \cite{salgado}), that is
\begin{equation}\label{4.3}
\tilde\nabla_j\de\psi/de{x^i}g^{ij}=m^2\psi\,.
\end{equation}
As standard, we take into account Gaussian normal coordinates
where $g_{4\alpha}=0$ and  $g_{44}=-1$. It is easy to show that
$\tilde\Gamma_{44}^{\;\;\;h}=0$ and then Eq. \eqref{4.3}, suitably
developed,  can be rewritten in the form
\begin{equation}\label{4.4}
\frac{\partial^2\psi}{\left(\partial x^4\right)^2} =
g^{\alpha\beta}\left(\frac{\partial^2\psi}{\partial x^\alpha
\partial x^\beta} -
\tilde\Gamma_{\alpha\beta}^{\;\;\;h}\de\psi/de{x^h}\right) -
m^2\psi\,,
\end{equation}
where
\begin{equation}\label{4.5}
\tilde\Gamma_{\alpha\beta}^{\;\;\;h}\de\psi/de{x^h}=
\frac{1}{2}g^{\lambda\gamma}\left(\de{g_{\alpha\gamma}}/de{x^\beta}
+ \de{g_{\beta\gamma}}/de{x^\alpha} -
\de{g_{\alpha\beta}}/de{x^\gamma}\right)\de\psi/de{x^\lambda} +
\frac{1}{2}\de{g_{\alpha\beta}}/de{x^4}\de\psi/de{x^4}\,.
\end{equation}
The Einstein-like equations are exactly of the form \eqref{3.5}
but now they depend also on  $\psi$
\begin{equation}\label{4.6}
\begin{split}
\tilde{R}_{ij} -\frac{1}{2}\tilde{R}g_{ij}= \frac{1}{\varphi}\Sigma_{ij} + \frac{1}{\varphi^2}\left( - \frac{3}{2}\de\varphi/de{x^i}\de\varphi/de{x^j} + \varphi\tilde{\nabla}_{j}\de\varphi/de{x^i} + \frac{3}{4}\de\varphi/de{x^h}\de\varphi/de{x^k}g^{hk}g_{ij} \right. \\
\left. - \varphi\tilde{\nabla}^h\de\varphi/de{x^h}g_{ij} - V\/(\varphi)g_{ij} \right)
\end{split}
\end{equation}
Adopting the notation of Sec.II, we can rewrite Eqs. \eqref{4.6}
in the form
\begin{equation}\label{4.7}
W_{ij}:=\varphi\tilde{G}_{ij} + T_{ij}=0\,,
\end{equation} where
\begin{equation}\label{4.8}
\begin{split}
-T_{ij}= \Sigma_{ij} + \frac{1}{\varphi}\left( - \frac{3}{2}\de\varphi/de{x^i}\de\varphi/de{x^j} + \varphi\tilde{\nabla}_{j}\de\varphi/de{x^i} + \frac{3}{4}\de\varphi/de{x^h}\de\varphi/de{x^k}g^{hk}g_{ij} \right.\\
\left. - \varphi\tilde{\nabla}^h\de\varphi/de{x^h}g_{ij} -
V\/(\varphi)g_{ij} \right)\,.
\end{split}
\end{equation}
We have now all the ingredients to discuss the Cauchy problem for
the system \eqref{4.3} and \eqref{4.7}. As first result, it has to
be pointed out that Eqs. \eqref{4.3} imply the conservation laws
\begin{equation}\label{4.9}
\tilde{\nabla}_j\Sigma^{ij}=0\,.
\end{equation}
In Appendix A, it is shown that Eqs. \eqref{4.9} are equivalent to
the conservation laws
\begin{equation}\label{4.10}
\tilde{\nabla}^jW_{ij}=0\,,
\end{equation}
see also \cite{Koivisto}. In this case, it is useful to remember
that both   metric-affine theories ({\it \`{a} la} Palatini and
with torsion) are  equivalent to  a metric scalar-tensor theory
with a Brans-Dicke parameter $\omega=-\frac{3}{2}$
\cite{CCSV1,Sotiriou,Sotiriou-Liberati1}.

For the the Cauchy problem, it is important to point out that the
second of Eqs. \eqref{A.5} is given by the  Klein-Gordon Eq.
\eqref{4.3}. Regarding the constraint  \eqref{A.6}, from the field
Eqs. \eqref{4.6}, it is easy to show that it involves only first
order derivatives with respect to the time variable $x^4$. In
order to show this point, let us observe that the r.h.s. of Eqs.
\eqref{4.6} contains the second derivatives of the scalar field
$\varphi=f'(F(\Sigma))$  and then implies the second derivatives
of the metric $g_{ij}$ and the third derivatives of the
Klein-Gordon field $\psi$.  In more details,   the second
derivatives of  $\varphi$ are present in the terms
\begin{subequations}\label{4.11}
\begin{equation}\label{4.11a}
\tilde\nabla_j\de\varphi/de{x^i}=
\frac{\partial^2\varphi}{\partial x^j \partial x^i} -
\tilde\Gamma_{ji}^{\;\;\;s}\de\varphi/de{x^s}\,,
\end{equation}
and
\begin{equation}\label{4.11b}
\tilde\nabla^h\de\varphi/de{x^h}g_{ij}=
\left(g^{hk}\frac{\partial^2\varphi}{\partial x^h \partial x^k} -
g^{hk}\tilde\Gamma_{hk}^{\;\;\;s}\de\varphi/de{x^s}\right)g_{ij}\,.
\end{equation}
\end{subequations}
Assuming as before Gaussian normal coordinates, which means
$g_{44}=g^{44}=-1$ and $g_{4\alpha}=g^{4\alpha}=0$, it is easy to
show that in the equations $\tilde{G}_{4i}=-T_{4i}$  there are no
second order time derivatives ${\displaystyle
\frac{\partial^2\varphi}{(\partial x^4)^2}}$, and the only time
derivatives are   ${\displaystyle \de\varphi/de{x^4}}$ and
${\displaystyle \frac{\partial^2\varphi}{\partial x^4
\partial x^\alpha}}$. On the other hand, it is useful to observe that these terms
involve also second order time derivatives of the Klein-Gordon
field ${\displaystyle \frac{\partial^2\psi}{(\partial x^4)^2}}$,
and ${\displaystyle \frac{\partial^3\psi}{\partial
x^\alpha(\partial x^4)^2}}$, the second order spatial derivatives
of the time derivative ${\displaystyle
\frac{\partial^3\psi}{\partial x^\alpha
\partial x^\beta \partial x^4}}$, and the derivatives of the metric
 ${\displaystyle \de{g_{\alpha\beta}}/de{x^4}}$ and ${\displaystyle \frac{\partial^2
g_{\alpha\beta}}{\partial x^\gamma \partial x^4}}$.

At this point, it is sufficient to observe that the Klein-Gordon
Eq. \eqref{4.4}, evaluated on the initial surface $x^4=0$, allows
to express the second order time derivative ${\displaystyle
\frac{\partial^2\psi}{(\partial x^4)^2}}$ and its spatial
derivatives  ${\displaystyle \frac{\partial^3\psi}{\partial
x^\alpha(\partial x^4)^2}}$ on the same surface  as a function of
the  Cauchy data ($g_{\alpha\beta}$, ${\displaystyle
\de{g_{\alpha\beta}}/de{x^4}}$, $\psi$, ${\displaystyle
\de\psi/de{x^4}}$) and of their spatial derivatives. The remaining
quantities ${\displaystyle \frac{\partial^3\psi}{\partial x^\alpha
\partial x^\beta \partial x^4}}$, ${\displaystyle \de{g_{\alpha\beta}}/de{x^4}}$,
and ${\displaystyle \frac{\partial^2 g_{\alpha\beta}}{\partial
x^\gamma
\partial x^4}}$, defined on $x^4=0$, can be directly calculated as
functions of the same Cauchy data.

In conclusion, the equations $\tilde{G}_{4i}=-T_{4i}$ on $x^4=0$
involve only the Cauchy data and their spatial derivatives. In
other words, they can be considered as genuine constraint
equations for the Cauchy data.

At this point,  some further remarks are necessary regarding the
corresponding Eqs. \eqref{A.5} for the case considered.  We have
already seen that Eqs. \eqref{4.6} imply second derivatives of the
metric $g_{ij}$ and third derivatives of the scalar field $\psi$.
Our task is now to prove that Eqs. \eqref{A.5} imply at most
second order time derivatives (i.e. with respect to the  variable
$x^4$) of the metric and at most first order time derivatives of
the Klein-Gordon field. The first request is obvious since only
second order derivatives of the metric are involved. Besides, the
second request is satisfied thanks to the Klein-Gordon Eq.
\eqref{4.4}. In fact, by deriving Eq. \eqref{4.4} with  respect to
the variable $x^4$ and with respect to the remaining variables
$x^\alpha$, by substituting Eq. \eqref{4.4} into the obtained
results and evaluating all the quantities on the initial surface
$x^4=0$, it is possible to express ${\displaystyle
\frac{\partial^3\psi}{(\partial x^4)^3}}$ and ${\displaystyle
\frac{\partial^3\psi}{\partial x^\alpha(\partial x^4)^2}}$ on $x^4
=0$ as functions of the Cauchy data and, at most, of the
derivatives ${\displaystyle \frac{\partial^2
g_{\alpha\beta}}{\partial(x^4)^2}}$.

In conclusion, in the equations corresponding to Eqs. \eqref{A.5},
only second order time derivatives of the metric appear.
Obviously, the initial value problem is well-formulated if these
equations are in normal form with respect to ${\displaystyle
\frac{\partial^2 g_{\alpha\beta}}{\partial(x^4)^2}}$ at least on
$x^4=0$. As in the perfect-fluid case, such a request could impose
further constraints on the initial data and, possibly, on the form
of the $f(R)$-function, but, in general, it works and the
corresponding Cauchy problem is well-formulated.

\section{Discussion and Conclusions}
In this paper, we have shown that the initial value problem for
metric-affine $f(R)$-gravity is  well-formulated. This means that
 there are no objections to the viability of metric--affine
$f(R)$-theories of gravity based on the well formulation of the
initial value problem. However, also the well--posed  problem is
necessary in order to achieve a complete control of dynamics but,
in this case, the role of source fields has to be carefully
discussed. This topic will be discussed in a furthcoming paper.
 
Since $f(R)$-gravity is a gauge theory, like General
Relativity, it is crucial the choice of suitable coordinates in
order to correctly formulate the problem. We have adopted Gaussian
normal coordinates which can be defined any time that the
derivative operator $\nabla_i$ arises from a metric $g_{ij}$. They
are also called {\it synchronous coordinates} and are particularly
useful for calculations on a given non-null surface $S_3$, i.e. an
3-dimensional embedded sub-manifold of the 4-dimensional manifold
$M$. They allow to define uniquely orthogonal geodesics to $S_3$
and then to correctly formulate the conditions of validity for the
Cauchy-Kowalewski theorem \cite{Wald}.

In metric-affine formalism, it is always possible to show that a
given $f(R)$-theory, in vacuo, can be recast in  Einstein's
gravity plus a cosmological constant. This means that the initial
value problem is always well-formulated and well-posed. The same
conclusion holds in the case of matter coupling  every time that
the trace of the stress-energy tensor is a constant.

As shown in \cite{CCSV1,CCSV2,CCSV3}, by introducing matter fields
in the Palatini and in the metric-affine approach with torsion,
one can define, in two steps $R=F(\Sigma)$ and
$\varphi:=f'(F(\Sigma)$, a suitable scalar field that allows: $i)$
to reduce the theory to scalar-tensor theory; $ii)$ to relate the
form of $f(R)$ to the trace of the matter (field) stress-energy
tensor. In this case, it is always possible a well-formulation of
the Cauchy problem avoiding the singularities which could emerge
with other gauge choices \cite{Faraoni}. Besides, matter fields
could induce further constraints on the Cauchy surface $x^4$
which, if suitably defined, lead to the normal form of the
matter-field equations. This is one of the main requests to obtain
a well-formulated value problem. However different fields, acting
as sources of the field equations,  like perfect fluids,
Yang-Mills, and Klein-Gordon  fields, could generate different
constraints on $x^4$.   Such constraints could imply also
restrictions on the possible forms of $f(R)$. In conclusion, as in
General Relativity, the gauge choice is essential for a correct
formulation of the initial value problem while the source fields
have to be carefully discussed to obtain a well-posed problem. In
a forthcoming paper, we will study specific problems and
$f(R)$-models where both the issues (i.e. well-formulation and
well-posedness) could be achieved.

\appendix
\section{The effective stress-energy tensor and the conservation laws}

\begin{Proposition}\label{ProB.1}
Eqs. \eqref{3.5}, \eqref{3.6} and \eqref{3.7} imply the usual
conservation laws $\tilde{\nabla}^j\/\Sigma_{ij}=0$
\end{Proposition}
\demo First of all, we recall that Eq. \eqref{3.6} and \eqref{3.7}
is equivalent to the relation
\begin{equation}\label{B.7bis}
\Sigma -\frac{6}{\varphi}V(\varphi) + 2V'(\varphi)=0
\end{equation}
(see \cite{CCSV1} for the proof). After that, taking the trace of
Eq. \eqref{3.5} into account, we get
\begin{equation}\label{B.8}
\Sigma= -\varphi\tilde{R} -
\frac{3}{2}\frac{1}{\varphi}\varphi_i\varphi^i +
3\tilde{\nabla}_i\varphi^i + \frac{4}{\varphi}V(\varphi)
\end{equation}
where for simplicity we have defined ${\displaystyle \varphi_i :=
\de\varphi/de{x^i}}$. Substituting Eq. \eqref{B.8} in Eq.
\eqref{B.7bis}, we obtain
\begin{equation}\label{B.9}
\tilde{R} + \frac{3}{2}\frac{1}{\varphi^2}\varphi_i\varphi^i -
\frac{3}{\varphi}\tilde{\nabla}_i\varphi^i +
\frac{2}{\varphi^2}V(\varphi) - \frac{2}{\varphi}V'(\varphi)=0
\end{equation}
We rewrite Eq. \eqref{3.5} in the form
\begin{equation}\label{B.10}
\begin{split}
\varphi\tilde{R}_{ij} -\frac{\varphi}{2}\tilde{R}g_{ij}= \Sigma_{ij} + \frac{1}{\varphi}\left( - \frac{3}{2}\varphi_i\varphi_j + \varphi\tilde{\nabla}_{j}\varphi_i + \frac{3}{4}\varphi_h\varphi^h\/g_{ij} + \right. \\
\left. - \varphi\tilde{\nabla}^h\varphi_h\/g_{ij} -
V\/(\varphi)g_{ij} \right)
\end{split}
\end{equation}
The covariant divergence of \eqref{B.10} yields
\begin{equation}\label{B.11}
\begin{split}
(\tilde\nabla^j\varphi)\tilde{R}_{ij} + \varphi\tilde\nabla^j\tilde{G}_{ij} -\frac{1}{2}\tilde{R}\tilde\nabla_i\varphi = \tilde\nabla^j\Sigma_{ij} + \left(\tilde\nabla^j\tilde{\nabla}_{j}\tilde\nabla_i - \tilde\nabla_i\tilde{\nabla}^j\tilde\nabla_j\right)\varphi +\\
+
\tilde\nabla^j\left[\frac{1}{\varphi}\left(-\frac{3}{2}\varphi_i\varphi_j
+ \frac{3}{4}\varphi_h\varphi^h\/g_{ij} -
V\/(\varphi)g_{ij}\right)\right]
\end{split}
\end{equation}
By definition, the Einstein and the Ricci tensors satisfy
$\tilde\nabla^j\tilde{G}_{ij}=0$ and
$(\tilde\nabla^j\varphi)\tilde{R}_{ij} =
\left(\tilde\nabla^j\tilde{\nabla}_{j}\tilde\nabla_i -
\tilde\nabla_i\tilde{\nabla}^j\tilde\nabla_j\right)\varphi$. Then
Eq. \eqref{B.11} reduces to
\begin{equation}\label{B.12}
-\frac{1}{2}\tilde{R}\tilde\nabla_i\varphi =
\tilde\nabla^j\Sigma_{ij}+
\tilde\nabla^j\left[\frac{1}{\varphi}\left(-\frac{3}{2}\varphi_i\varphi_j
+ \frac{3}{4}\varphi_h\varphi^h\/g_{ij} -
V\/(\varphi)g_{ij}\right)\right]
\end{equation}
Finally, making use of Eq. \eqref{B.9} it is easily seen that
\begin{equation}\label{B.13}
-\frac{1}{2}\tilde{R}\tilde\nabla_i\varphi =
\tilde\nabla^j\left[\frac{1}{\varphi}\left(-\frac{3}{2}\varphi_i\varphi_j
+ \frac{3}{4}\varphi_h\varphi^h\/g_{ij} -
V\/(\varphi)g_{ij}\right)\right]
\end{equation}
from which the conclusion $\tilde\nabla^j\Sigma_{ij}=0$ follows
\enddemo

\begin{Proposition}\label{ProB.2}
Given the Levi-Civita connection
\begin{equation}\label{B.1}
\bar{\Gamma}_{ij}^{\;\;\;h}= \tilde{\Gamma}_{ij}^{\;\;\;h} +
\frac{1}{2\varphi}\de\varphi/de{x^j}\delta^h_i -
\frac{1}{2\varphi}\de\varphi/de{x^p}g^{ph}g_{ij} +
\frac{1}{2\varphi}\de\varphi/de{x^i}\delta^h_j
\end{equation}
associated with the conformal metric tensor $\bar{g}=\varphi\/g$,
and given the effective energy--impulse tensor
\begin{equation}\label{B.2}
T_{ij}= \frac{1}{\varphi}\Sigma_{ij} -
\frac{1}{\varphi^3}V\/(\varphi)\bar{g}_{ij}
\end{equation}
the condition $\bar{\nabla}^j\/T_{ij}=0\/$ is equivalent to the
condition $\tilde{\nabla}^j\/\Sigma_{ij}=0$
\end{Proposition}
\demo
\begin{equation}\label{B.3}
\begin{split}
\bar{\nabla}^j\/T_{ij}= \frac{1}{\varphi}g^{sj}\bar{\nabla}_s\/T_{ij}= \frac{1}{\varphi}g^{sj}\left[\tilde{\nabla}_s\/T_{ij} - \frac{1}{2\varphi}\left(\de{\varphi}/de{x^i}\delta^q_s + \de{\varphi}/de{x^s}\delta^q_i - \de{\varphi}/de{x^u}g^{uq}g_{si}\right)T_{qj} +\right. \\
\left. - \frac{1}{2\varphi}\left(\de{\varphi}/de{x^j}\delta^q_s +
\de{\varphi}/de{x^s}\delta^q_j -
\de{\varphi}/de{x^u}g^{uq}g_{sj}\right)T_{iq}\right]
\end{split}
\end{equation}
We have separately
\begin{equation}\label{B.4}
\begin{split}
\frac{1}{\varphi}g^{sj}\tilde{\nabla}_s\/T_{ij}=\frac{1}{\varphi}g^{sj}\tilde{\nabla}_s\/\left(\frac{1}{\varphi}\Sigma_{ij} - \frac{1}{\varphi^2}V\/(\varphi)g_{ij}\right)=\frac{1}{\varphi^2}\tilde{\nabla}^j\Sigma_{ij} - \frac{1}{\varphi^3}\de{\varphi}/de{x^s}\Sigma_i^{\;s} +\\
- \frac{1}{\varphi}\de
/de{x^s}\left(\frac{1}{\varphi^2}V\/(\varphi)\right)\delta^s_i
\end{split}
\end{equation}
\begin{equation}\label{B.5}
\begin{split}
\frac{1}{\varphi}g^{sj}\frac{1}{2\varphi}\left(\de{\varphi}/de{x^i}\delta^q_s + \de{\varphi}/de{x^s}\delta^q_i - \de{\varphi}/de{x^u}g^{uq}g_{si}\right)T_{qj}=\\
=\frac{1}{\varphi}g^{sj}\frac{1}{2\varphi}\left(\de{\varphi}/de{x^i}\delta^q_s + \de{\varphi}/de{x^s}\delta^q_i - \de{\varphi}/de{x^u}g^{uq}g_{si}\right)\/\left(\frac{1}{\varphi}\Sigma_{qj} - \frac{1}{\varphi^2}V\/(\varphi)g_{qj}\right)=\\
=\frac{1}{2\varphi^3}g^{sj}\/\left(\de{\varphi}/de{x^i}\Sigma_{sj} + \de{\varphi}/de{x^s}\Sigma_{ij} - \de{\varphi}/de{x^u}g_{si}\Sigma^u_{\;j}\right)+\\
-\frac{1}{2\varphi^4}g^{sj}\/\left(\de{\varphi}/de{x^i}V\/(\varphi)g_{sj} + \de{\varphi}/de{x^s}V\/(\varphi)g_{ij} - \de{\varphi}/de{x^u}V\/(\varphi)\delta^u_jg_{si}\right)=\\
=\frac{1}{2\varphi^3}\de{\varphi}/de{x^i}\Sigma -
\frac{2}{\varphi^4}\de{\varphi}/de{x^i}V\/(\varphi)
\end{split}
\end{equation}
\begin{equation}\label{B.6}
\begin{split}
\frac{1}{\varphi}g^{sj}\frac{1}{2\varphi}\left(\de{\varphi}/de{x^j}\delta^q_s + \de{\varphi}/de{x^s}\delta^q_j - \de{\varphi}/de{x^u}g^{uq}g_{sj}\right)T_{iq}=\\
=\frac{1}{\varphi}g^{sj}\frac{1}{2\varphi}\left(\de{\varphi}/de{x^j}\delta^q_s + \de{\varphi}/de{x^s}\delta^q_j - \de{\varphi}/de{x^u}g^{uq}g_{sj}\right)\/\left(\frac{1}{\varphi}\Sigma_{iq} - \frac{1}{\varphi^2}V\/(\varphi)g_{iq}\right)=\\
=\frac{1}{2\varphi^3}g^{sj}\/\left(\de{\varphi}/de{x^j}\Sigma_{is} + \de{\varphi}/de{x^s}\Sigma_{ij} - \de{\varphi}/de{x^u}g_{sj}\Sigma^u_{\;i}\right)+\\
-\frac{1}{2\varphi^4}g^{sj}\/\left(\de{\varphi}/de{x^j}V\/(\varphi)g_{si} + \de{\varphi}/de{x^s}V\/(\varphi)g_{ij} - \de{\varphi}/de{x^u}V\/(\varphi)\delta^u_ig_{sj}\right)=\\
-\frac{1}{\varphi^3}\de{\varphi}/de{x^s}\Sigma^{\;s}_i +
\frac{1}{\varphi^4}\de{\varphi}/de{x^i}V\/(\varphi)
\end{split}
\end{equation}
Collecting Eqs. \eqref{B.4}, \eqref{B.5} and \eqref{B.6} we have
then
\begin{equation}\label{B.7}
\bar{\nabla}^j\/T_{ij}=\frac{1}{\varphi^2}\tilde{\nabla}^j\Sigma_{ij}
+ \frac{1}{\varphi^3}\de{\varphi}/de{x^i}\left[-\frac{1}{2}\Sigma +
\frac{3}{\varphi}V(\varphi) - V'(\varphi)\right]=
\frac{1}{\varphi^2}\tilde{\nabla}^j\Sigma_{ij}
\end{equation}
because the identity $-\frac{1}{2}\Sigma +
\frac{3}{\varphi}V(\varphi) - V'(\varphi)=0$ holds identically,
being equivalent to the definition $\varphi=f'(F(\Sigma))$
\cite{CCSV1}.
\enddemo

\end{document}